\documentstyle[]{mn}

\title[The kilo-second QPOs of GK~Per]{Solving the kilo-second QPO
   problem of the intermediate polar GK~Persei}

\author[L. Morales-Rueda, M.\,D.\ Still, P. Roche]
       {L. Morales-Rueda$^{1}$, M.\,D.\ Still$^{2}$ and P. Roche$^{1}$\\ 
       $^{1}$Astronomy Centre, University of Sussex, Falmer,
       Brighton BN1 9QJ (lmorales@star.cpes.susx.ac.uk, 
       pdr@star.cpes.susx.ac.uk)\\
       $^{2}$Physics and Astronomy, University of St. Andrews , North
       Haugh, St. Andrews, Fife KY16 9SS (mds1@st-and.ac.uk)}

\date{Accepted 1999 January 20. Received 1998 December 16; in original
  form 1998 July 29.}

\pagerange{\pageref{firstpage}--\pageref{lastpage}}

\def\LaTeX{L\kern-.36em\raise.3ex\hbox{a}\kern-.15em
    T\kern-.1667em\lower.7ex\hbox{E}\kern-.125emX}

\begin{document}

\newcommand{\gk}{\mbox{GK~Per}} 
\newcommand{\etal}{\mbox{et\ al.\ }}
\newcommand{\kmsec}{\,\mbox{$\mbox{km}\,\mbox{s}^{-1}$}}
\newcommand{\phispin}{$\phi_{spin}$}
\newcommand{\hb}{\hbox{$\hbox{H}\beta$}}
\newcommand{\hgam}{\hbox{$\hbox{H}\gamma$}}
\newcommand{\heii}{\hbox{$\hbox{He\,{\sc ii}\,$\lambda$4686\,\AA}$}}
\newcommand{\ciiiniii}{\hbox{$\hbox{C\,{\sc iii}/N\,{\sc
        iii}\,$\lambda\lambda$4640--50\,\AA}$}}

\label{firstpage}

\maketitle

\begin{abstract}
  
  We detect the likely optical counterpart to previously reported
  X-ray QPOs in spectrophotometry of the intermediate polar \gk\ 
  during the 1996 dwarf nova outburst.  The characteristic timescales
  range between 4000--6000\,s. Although the QPOs are an order of
  magnitude longer than those detected in the other dwarf novae we
  show that a new QPO model is not required to explain the long
  timescale observed.  We demonstrate that the observations are
  consistent with oscillations being the result of normal-timescale
  QPOs beating with the spin period of the white dwarf. We determine
  the spectral class of the companion to be consistent with its
  quiescent classification and find no significant evidence for
  irradiation over its inner face.  We detect the white dwarf spin
  period in line fluxes, V/R ratios and Doppler-broadened emission
  profiles.

\end{abstract}


\section{Introduction}

\gk\ (Nova Per 1901; Campbell 1903), belongs to a subgroup of
cataclysmic variables (CVs) called Intermediate Polars (IPs). In these
systems an asynchronously-rotating, magnetic white dwarf accretes
material from a less-massive, late-type companion filling its Roche
lobe.  Gas leaving the companion star attempts to form an accretion
disc around the primary star but its magnetic field either prevents
the formation of the disc or truncates it near the white dwarf.

\gk\ was identified with the X-ray source A0327+43 by King, Ricketts
\& Warwick (1979) and confirmed as an IP by the detection of a 351\,s
X-ray spin pulse by Watson, King \& Osborne (1985; hereafter WKO) and
Norton, Watson \& King (1988).  The same period was subsequently found
in optical photometry by Patterson (1991).  \gk\ has the longest
orbital period from the sample of known CVs, P$_{\mbox{orb}}$ = 2\,d,
(Crampton, Cowley \& Fisher 1986; hereafter CCF).  The wide binary
separation combined with a relatively weak magnetic field ($\sim$~1
MG) means that a truncated accretion disc must be present if current
theories of disc formation are correct (Hameury, King \& Lasota 1986).
The presence of a disc has yet to be confirmed by direct observation,
although the system does undergo dwarf nova outbursts every
2--3\,years where its optical brightness increases from 13th to 10th
magnitude (Sabbadin \& Bianchini 1983).  The most-likely mechanism for
dwarf nova outbursts is a thermal instability within an accretion disc
(Osaki 1974).  \gk\ outbursts have been modelled as such by Cannizzo
\& Kenyon (1986) and Kim, Wheeler \& Mineshige (1992).

This paper is a continuation of paper {\sc i} (Morales-Rueda, Still \&
Roche 1996), in which we presented spectrophotometric observations of
\gk\ taken on the rise to its 1996 outburst (Mattei \etal 1996). We
reported the detection of quasi-periodic oscillations (QPOs) within
the Doppler-broadened emission lines of H {\sc i} and He {\sc ii}.
This provides an opportunity to map the velocity structure of the
oscillations.  QPOs are defined as low-coherence brightness
oscillations thought to be associated with material within the inner
accretion flows of CVs.  Theoretical models developed to explain QPOs
consider the presence of dense blobs of material orbiting in the inner
regions of the accretion disc (Bath 1973), or non-radial pulsations
over the surface of the white dwarf (Papaloizou \& Pringle 1978), or
radially-oscillating acoustic waves in the inner disc (Okuda \etal
1992; Godon 1995).  In these models, the QPO timescales match
observations of dwarf novae and are of the order of a few hundred
seconds.  However the QPO periods detected in \gk\ are an order of
magnitude longer than this. Previous to this paper they have only been
detected in X-ray data taken during outbursts; WKO discovered them in
1.5--8.5 keV EXOSAT data at the peak of the 1983 outburst, while
Ishida \etal (1996) report a second detection at 0.7--10\,keV with
ASCA during the rise to the 1996 outburst discussed in this paper.

To explain the long timescales WKO suggested the QPO mechanism is
caused by beating between the 351\,s white dwarf spin period and
inhomogeneous gas orbiting at the inner edge of the accretion disc.
Hellier \& Livio (1994; hereafter HL) noted that the X-ray hardness
ratio varies over the QPO cycle as expected from photoelectric
absorption by cool gas and that a period of a few thousand seconds is
consistent with the orbital frequency of gas if it is deposited onto
the disc by a gas stream which has partially avoided impacting the
outer disc rim and follows a ballistic trajectory.  They propose that
the QPO mechanism is X-ray absorption by vertically-extended blobs of
gas orbiting at this preferred inner impact radius. In paper {\sc i}
we determined that the characteristic velocity structure of the
optical counterpart to the QPOs observed by Ishida \etal (1996) is
consistent with blobs in the inner disc. In the current paper we
present further analysis which indicates that the optical QPO is also
driven by absorption, but favours strongly a beat model over the
disc-overflow interpretation.

\section{Observations}

\begin{figure}
\begin{picture}(100,0)(10,20)
\put(0,0){\includegraphics{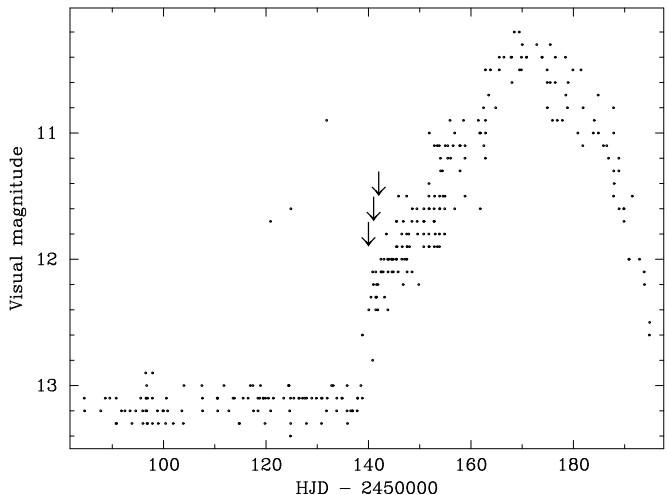}}
\noindent
\end{picture}
\vspace{75mm}
\caption{The visual light curve of \gk\ during the 1996 outburst
  obtained from the Variable Star Network
  (http://www.kusastro.kyoto-u.ac.jp/vsnet/). 
  The arrows indicate the times of our spectrophotometric
  observations.}
\end{figure}

Between 1996 February 26 and 28, 6--8 days before the ASCA pointings
of Ishida \etal (1996), we obtained spectrophotometry of \gk\ using
the Intermediate Dispersion Spectrograph mounted on the 2.5\,m Isaac
Newton Telescope (INT) on La Palma. Table~1 gives a journal of
observations.  In Fig.~1 we show a visual light curve obtained by the
Variable Star Network during the 1996 outburst, with arrows indicating
the days on which we made observations.  The quick readout mode was
used in conjunction with a Tektronix CCD windowed to
1024\,$\times$\,150 pixels to reduce dead time and obtain good
sampling of the spin cycle.  The exposure times and resolution of the
data were already described in paper {\sc i}.

After debiasing and flat-fielding the frames by tungsten lamp
exposures, spectral extraction proceeded according to the optimal
algorithm of Horne (1986).  The data were wavelength calibrated using
a CuAr arc lamp and corrected for instrumental response and extinction
using the flux standard HZ\,15 (Stone 1977).  The spectrograph slit
orientation of PA~$249.1^{\circ}$ allowed a 15th magnitude nearby star
approximately 0.5\,arcsec ENE of \gk\ to be employed as calibration
for light losses on the slit.

We also have available to us spectroscopy of various K-type stars from
1995 October 11 to 13 obtained from the INT and from 1995 October 30
to November 2 with the 2.1\,m telescope in the McDonald Observatory in
Texas. The INT instrumental setup was identical to the one used for
the 1996 observations described above.  For the McDonald data, the
low-to-moderate resolution spectrometer ES2 was employed in
conjunction with the TI1 CCD and a grating ruled at 1200
lines\,mm$^{-1}$ covering the wavelength region $\lambda$4196 \AA
--$\lambda$4894 \AA\ giving a resolution of 200 km~s$^{-1}$ at \hb.

The spectra were flat-fielded, optimally extracted and wavelength
calibrated also in the standard manner. Flux calibrations were applied
using observations of the standards HD19445 (Oke \& Gunn 1983) and
Feige~110 (Stone 1977) for the October and November data respectively.
Table~2 gives a list of the K-type templates observed over both runs.

\begin{table}
\centering
\begin{minipage}{84mm}
\caption{Journal of observations. $E$ is the cycle number plus binary 
  phase with respect to the ephemeris given by Crampton, Cowley \&
  Fisher (1986). Phases have been adjusted by $\pi/2$ so that phase 0
  corresponds to superior conjunction of the white dwarf.}
\begin{center}
\begin{tabular}{cccccr}
\multicolumn{1}{c}{Date} & \multicolumn{1}{c}{Start} & 
\multicolumn{1}{c}{End} & \multicolumn{1}{c}{Start} & 
\multicolumn{1}{c}{End} & \multicolumn{1}{c}{No. of} \\
\multicolumn{1}{c}{(1996 Feb)} & \multicolumn{2}{c}{(UT)} & 
\multicolumn{2}{c}{($E -$ 2\,000)} & \multicolumn{1}{c}{spectra} \\ \hline
26&20.06 &0.12 &617.119 &617.204 &109 \\
27&20.04 &0.07 &617.620 &617.704 &116 \\
28&20.06 &0.04 &618.121 &618.204 &117 \\
\end{tabular}
\end{center}
\end{minipage}
\end{table}

\section{Results}

\subsection{Average spectra}

\begin{figure*}
\begin{picture}(100,0)(10,20)
\put(0,0){\includegraphics{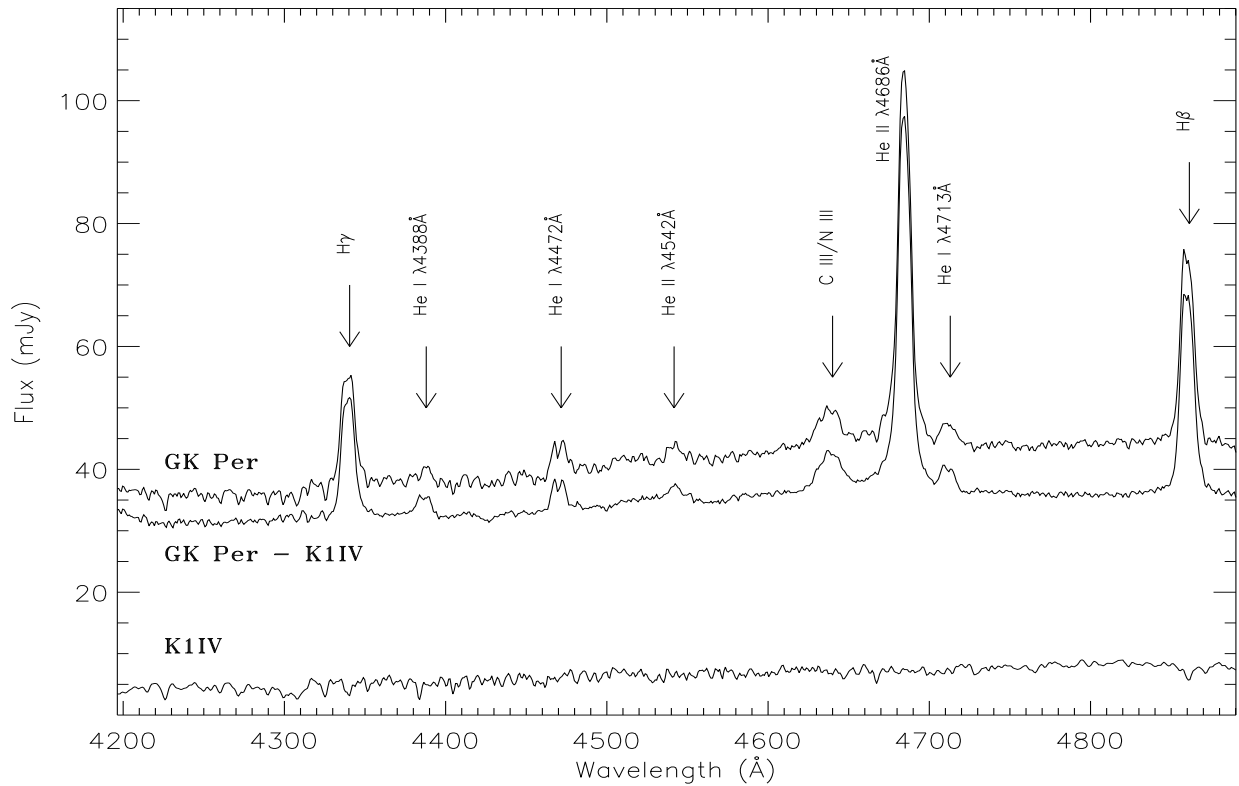}}
\noindent
\end{picture}
\vspace{120mm}
\caption{The top spectrum is an average of all \gk\ spectra
  obtained during the third night of observations. The bottom spectrum
  corresponds to the K1IV template HD197964 multiplied by 3$\times
  10^{-4}$.  The middle spectrum is the residual resulting from the
  subtraction of the template from the averaged \gk\ spectrum and
  probably resembles the spectrum of the accretion flow.}
\end{figure*}

Fig.~2 presents the average of all the data collected on 1996 Feb 28.
It is characterised by a flat continuum, broad Balmer and He\,{\sc i}
lines in emission, high excitation lines of He\,{\sc ii}, N\,{\sc
  iii}\, and C\,{\sc iii} and numerous faint, narrow absorption
features of Fe\, {\sc i}, Ca\,{\sc i}, Ti\,{\sc ii} and Sr\,{\sc ii}
that had been identified as signatures of the K-type secondary star by
Kraft (1964), Gallagher \& Oinas (1974), CCF and Reinsch (1994).

We employed K star spectral templates to determine which luminosity
class best matched the secondary star in this system during outburst
and search for signatures of increased X-ray irradiation. Using CCF's
fit to the orbital radial velocity of the secondary star we shifted
out the orbital motion of the absorption lines with a quadratic
rebinning algorithm. We binned in velocity the spectra of \gk\ and the
K-type templates to ensure that they all had identical wavelength
ranges and dispersions.  We employed the optimal subtraction algorithm
of Marsh, Robinson \& Wood (1994) to determine the K star spectral
type -- we multiply the template by a monochromatic constant which
represents the contribution to the spectrum from non-stellar sources
of light and subtract the resulting spectrum from the \gk\ data.  The
residual was smoothed using a high-pass band filter (FWHM of gaussian
= 13 \AA), and a $\chi^2$ test performed between the original and
smoothed residual. This is an iterative procedure to determine the
optimum value of the monochromatic constant which continues until
$\chi^2$ is minimised. Table~2 lists the templates, their spectral
classes, and the reduced $\chi^2$ obtained after applying optimal
subtraction.  The best fit template is the K1{\sc iv} star HD197964
which provided a reduced $\chi^2$ of 2.5. The secondary star
contributes 13 per cent of the total light in this spectral region on
the third night of observations. This compares to 33 per cent found by
CCF and Gallagher \& Oinas (1974) during quiescence indicating that
the accretion flow has increased in brightness.

\begin{table}
\centering
\begin{minipage}{84mm}
\caption{A list of template K star used to determine the best-fit spectral
type for the secondary star.}
\begin{center}
\begin{tabular}{llllll}
\multicolumn{1}{l}{Name} & \multicolumn{1}{l}{Spectral} &
\multicolumn{1}{l}{$\chi^2$} & 
\multicolumn{1}{l}{Name} & \multicolumn{1}{l}{Spectral} &
\multicolumn{1}{l}{$\chi^2$}\\
& \multicolumn{1}{l}{type} & & & \multicolumn{1}{l}{type} & \\
\hline
\multicolumn{3}{c}{October 1995} & \multicolumn{3}{c}{November 1995} \\
\hline

HR190  & K1{\sc iii} & 4.3 & 13 Lac & K0{\sc iii} & 4.0\\
HR8688 & K1{\sc iii} & 6.6 & 1 Peg & K1{\sc iii} & 5.7\\
HR8415 & K2{\sc iii} & 6.5 & 69 Aql & K2{\sc iii} & 3.9\\
HR8632 & K3{\sc iii} & 6.2 & 39 Cyg & K3{\sc iii} & 4.0\\
HR8974 & K1{\sc iv}  & 5.9 &33 Vul & K3.5{\sc iii} & 3.8\\
HR8881 & K1{\sc v} & 6.2 & 3$\eta$Cep & K0{\sc iv} & 5.9\\
HR222  & K2{\sc v} & 3.0 &HD197964 & K1{\sc iv} & 2.5\\
HR8832 & K3{\sc v} & 3.2 & & & \\

\end{tabular}
\end{center}
\end{minipage}
\end{table}

The best-fit luminosity classes are consistent with the quiescent
classifications of K2{\sc iv}p by Kraft (1964), K2{\sc iv}p--K2{\sc v}
by Gallagher \&\ Oinas (1974), K0{\sc iii} by CCF, and K3{\sc v} by
Reinsch (1994).  We find the spectral type to be constant across our
two phase samples - one during which a large area of the white
dwarf-facing surface is visible and the other when it is mostly
limb-occulted. Consequently there is no observational evidence for an
increase in irradiating flux from the accretion regions over the inner
face of the companion star, although we are limited by a small range
of spectral templates and poor orbital sampling.

\begin{table*}
\caption{Emission line fluxes in units of 10$^{-13}$ erg cm$^{-2}$ s$^{-1}$ 
from the spectra of  \gk\ on 1996   Feb 26-28. The minimum  and maximum
fluxes throughout the  night as well  as the average nightly  flux are
provided.     The  error   on  the   average   flux   is  the standard
deviation. Negative fluxes correspond to line absorption.}
\begin{center}
\begin{tabular}{lcccccc}
\multicolumn{1}{l}{Line} & \multicolumn{1}{c}{Flux} &
\multicolumn{1}{c}{Average} & \multicolumn{1}{c}{Flux} &
\multicolumn{1}{c}{Average} & \multicolumn{1}{c}{Flux} &
\multicolumn{1}{c}{Average}\\
&\multicolumn{1}{c}{range} & \multicolumn{1}{c}{flux} &
\multicolumn{1}{c}{range} & \multicolumn{1}{c}{flux} &
\multicolumn{1}{c}{range} & \multicolumn{1}{c}{flux}\\
\hline
& \multicolumn{2}{c}{26/2/96} & \multicolumn{2}{c}{27/2/96} &
\multicolumn{2}{c}{28/2/96}\\
\hline

H$\gamma$ &1.24 -- 2.97&2.05 $\pm$ 0.34&1.57 -- 3.31&2.27 $\pm$ 0.30&
2.09 -- 4.03&3.06 $\pm$ 0.37\\
H$\beta$ &1.43 -- 3.63&2.32 $\pm$ 0.40&1.73 -- 3.27&2.45 $\pm$
0.35&2.17 -- 4.57&3.41 $\pm$ 0.46\\
He\,{\sc i}\, $\lambda$4387.9\AA &-0.70 -- 0.31&-0.04 $\pm$ 0.15&-0.18
-- 0.29&0.03 $\pm$ 0.08&-0.14 -- 0.35&0.11 $\pm$ 0.09\\
He\,{\sc i}\, $\lambda$4437.6\AA &-0.83 -- 036& -0.29 $\pm$ 0.18&-0.60
-- -0.15&-0.34 $\pm$ 0.09&-0.68 -- -0.23&-0.44 $\pm$ 0.09\\
He\,{\sc i}\, $\lambda$4471.7\AA &-0.47 -- 0.72&0.13 $\pm$ 0.18&-0.36
-- 0.41&0.12 $\pm$ 0.14&-0.28 -- 0.68&0.25 $\pm$ 0.18\\
He\,{\sc i}\, $\lambda$4713.2\AA &-0.70 -- 0.52&0.02 $\pm$ 0.14&-0.18
-- 0.27&0.05 $\pm$ 0.08&-0.22 -- 0.31&0.02 $\pm$ 0.10\\
He\,{\sc i}\, $\lambda$4921.9\AA &-0.06 -- 0.93&0.38 $\pm$ 0.15&0.18
-- 0.58&0.39 $\pm$ 0.09&0.26 -- 1.03&0.68 $\pm$ 0.13\\
He\,{\sc ii}\, $\lambda$4541.7\AA &-0.46 -- 0.25&-0.06 $\pm$
0.14&-0.30 -- 0.30&0.02 $\pm$ 0.10&-0.21 -- 0.26&0.02 $\pm$ 0.11\\
He\,{\sc ii}\, $\lambda$4685.8\AA &4.01 -- 7.36&5.23 $\pm$ 0.77&3.24
-- 7.72&5.44 $\pm$ 0.78&6.04 -- 12.19&8.27 $\pm$ 0.94\\
Bowen blend &0.04 -- 1.05&0.50 $\pm$ 0.17&0.39 -- 1.08&0.65 $\pm$
0..13&0.71 -- 1.89&1.06 $\pm$ 0.17\\

\end{tabular}
\end{center}
\end{table*}

In order to measure integrated emission line fluxes from each of the
three nights, we fitted a third order polynomial through wavelength
bands relatively free of line features ($\lambda\lambda$4147-4212\AA,
$\lambda\lambda$4278-4306\AA, $\lambda\lambda$4560-4608\AA,
$\lambda\lambda$4770-4838\AA\ and subtracted the fit from the data.
Fluxes were measured by summing under each line profile and these are
provided in Table~3.  The intensity of the continuum, the lines and
the relative intensity of \heii\ with respect to the Balmer lines,
increases from the first night to the last as the system approaches
the outburst maximum.  We fit the emission lines during the three
nights with a power law function of time $F \sim t^\alpha$, and
provide the index $\alpha$ for each line in Table~4.

Power-law fits of the form $f_{\nu}=\nu^{\alpha}$ on each consecutive
night provide $\alpha$ = $-$1.61 $\pm$ 0.03, $-$1.43 $\pm$ 0.03 and
$-$1.39 $\pm$ 0.11.  Continuum slope changes slightly within
statistical uncertainties during the observing run, the spectra
becoming bluer with time consistent with a rise in temperature through
the accretion flow.  These indices are inconsistent with an accretion
disc emitting as a discrete set of blackbodies (Pringle 1981).

A comparison of the Feb 28 averaged spectrum with the spectra
presented by Reinsch (1994) reveals that the Balmer line fluxes are
$\sim$1.7 times larger than during quiescence and the \heii\ feature
and the \ciiiniii\ Bowen blend are $\sim$5.3 times brighter. He\,{\sc
  i}\, $\lambda$4471.7\AA\ and He\,{\sc i}\, $\lambda$4921.9\AA\ are
1.4 and 2.3 times brighter during this outburst stage, respectively.
In quiescence the Balmer lines are the brightest emission lines,
whereas the strongest line in the current data is \heii.  Szkody,
Mattei \& Mateo (1985) and CCF present spectra of \gk\ taken during
the 1983 outburst maximum and 20 days after outburst respectively in
which this behaviour is also clear.

\begin{table}
\centering
\begin{minipage}{84mm}
\caption{Power indices $\alpha$ obtained from fitting with a power law
  function to the integrated flux of each emission line and the
  continuum over the three nights of observations.}
\begin{center}
\begin{tabular}{lclc}
Line&Power index&Line&Power index\\
\hline
\hgam&2.3 $\pm$ 0.1&\hb&2.7 $\pm$ 0.1\\
He\,{\sc i}\,$\lambda4388$\AA&1.1 $\pm$ 0.4&He\,{\sc
  i}\,$\lambda4438$\AA&2.5 $\pm$ 0.8\\
He\,{\sc i}\,$\lambda4472$\AA&2.1 $\pm$ 0.7&He\,{\sc
  i}\,$\lambda4922$\AA&3.5 $\pm$ 0.5\\
\heii&3.5 $\pm$ 0.1&Bowen blend&2.3 $\pm$ 0.2\\
Continuum&2.41 $\pm$ 0.01& &\\
\end{tabular}
\end{center}
\end{minipage}
\end{table}

\subsection{Radial velocities}

In paper {\sc i} we provided an analysis of the emission line
velocities.  To complete the radial velocity analysis we now consider
the absorption lines.  In Sec.~3.1 we determined that our best
secondary star template has a spectral type of K1{\sc iv}. By masking
out the emission lines in individual \gk\ data and subtracting fits to
the continua from all spectra, we were able to cross-correlate the
absorption spectrum of \gk\ with our template (Tonry \& Davis 1979).
We corrected the resulting radial velocities by the systemic velocity
of the template star (-6.5\kmsec; Evans 1979) and fitted them with a
circular function:
\begin{equation}
V = \gamma + K \sin{2 \pi \left[ \phi - \phi_0\right]}
\end{equation}
Orbital phases were adopted relative to the corrected CCF ephemeris,
where $\phi_0$ corresponds to superior conjunction of the white dwarf.
$\gamma$ represents the systemic velocity of the binary, $K$ is the
radial velocity semiamplitude of the companion star and $\phi$ is the
orbital phase.

We combined the radial velocities measured by previous authors (Kraft
1964; CCF; Reinsch 1994) with our own values and plot them together in
Fig.~3. We assume that the errors on all individual measurements
previous to this study are equal to the mean error of 20 \kmsec.
The solid curve is the fit to all data, providing $\gamma = 30 \pm
1$\kmsec, $K = 119 \pm 2$\kmsec and $\phi_0 = 0.998 \pm 0.003$.  The
dot-dashed curve is a fit to all the data excluding the current set
where $\gamma = 22 \pm 2$\kmsec, $K = 128 \pm 2$\kmsec and $\phi_0 =
0.009 \pm 0.003$, providing reasonable agreement although the fits are
not consistent within the given errors. Martin (1988) showed that an
elliptical fit can account approximately for irradiation processes
over the inner face of the secondary star. Elliptical fits to the
quiescent data have already been produced by CCF and Reinsch (1994).
We do not have suitable phase sampling to produce a significant
elliptical fit with the current data. Therefore although we find no
evidence for secondary star irradiation in the absorption line radial
velocities during outburst, our phase coverage prevents us from ruling
it out.

\begin{figure*}
\begin{picture}(100,0)(10,20)
\put(0,0){\includegraphics{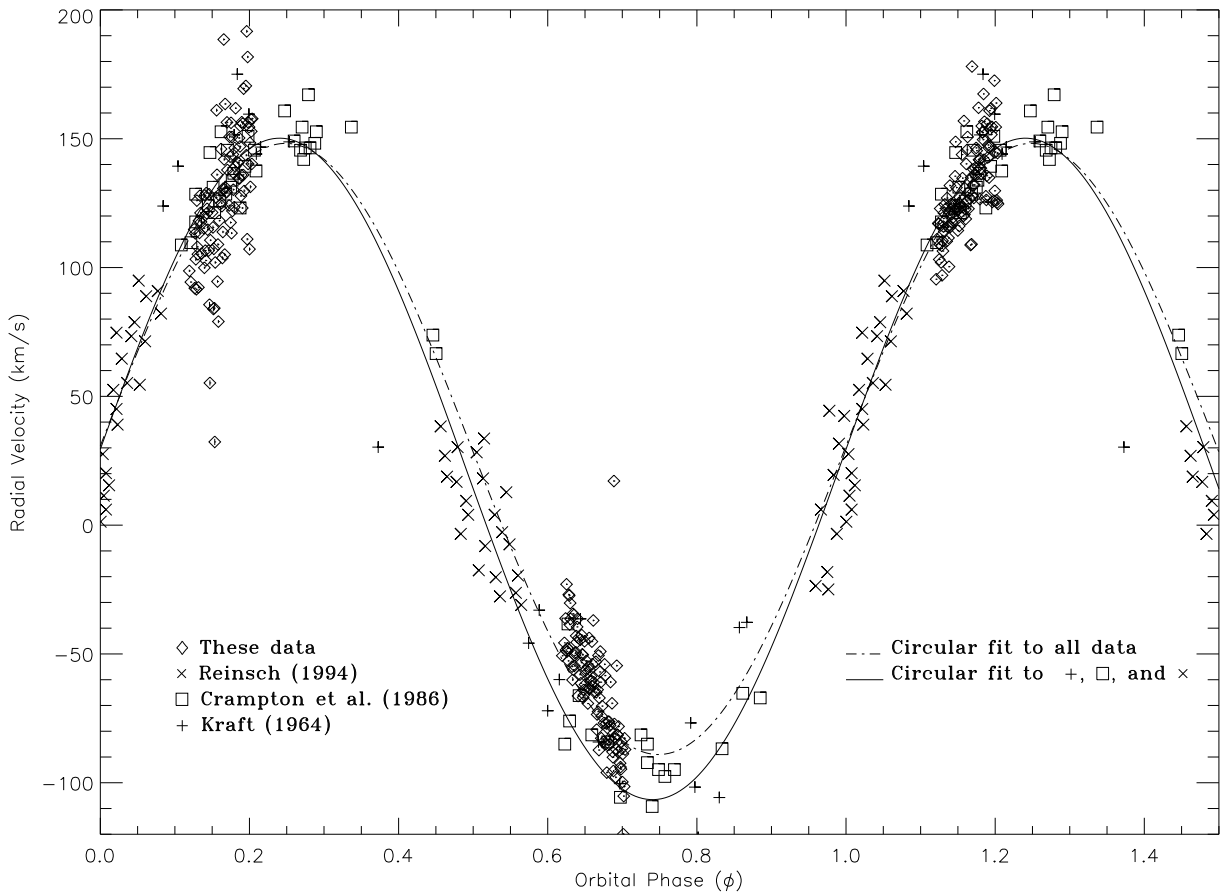}}
\noindent
\end{picture}
\vspace{140mm}
\caption{Radial velocities obtained from the current data by
  cross-correlating absorption features from the companion star
  against a K1{\sc iv} template, combined with similar measurements
  during quiescence from Kraft (1964), CCF, and Reinsch (1994). We
  provide two circular fits to the data.}
\end{figure*}

\subsection{Emission line profiles}

\begin{figure*}
\begin{picture}(100,0)(10,20)
\put(0,0){\includegraphics{fig4.eps}}
\noindent
\end{picture}
\vspace{140mm}
\caption{Trailed spectra of \hb, \heii\ and the average of He\,{\sc
    i}\,$\lambda4388$\AA, He\,{\sc i}\,$\lambda4472$\AA, He\,{\sc
    i}\,$\lambda4713$\AA, He\,{\sc i}\,$\lambda4922$\AA). Each row
  corresponds to a different night, the top row being the first night
  of observations. Images have separated linear greyscales where black
  corresponds to emission.}
\end{figure*}

\begin{figure*}
\begin{picture}(100,0)(10,20)
 \put(0,0){\includegraphics{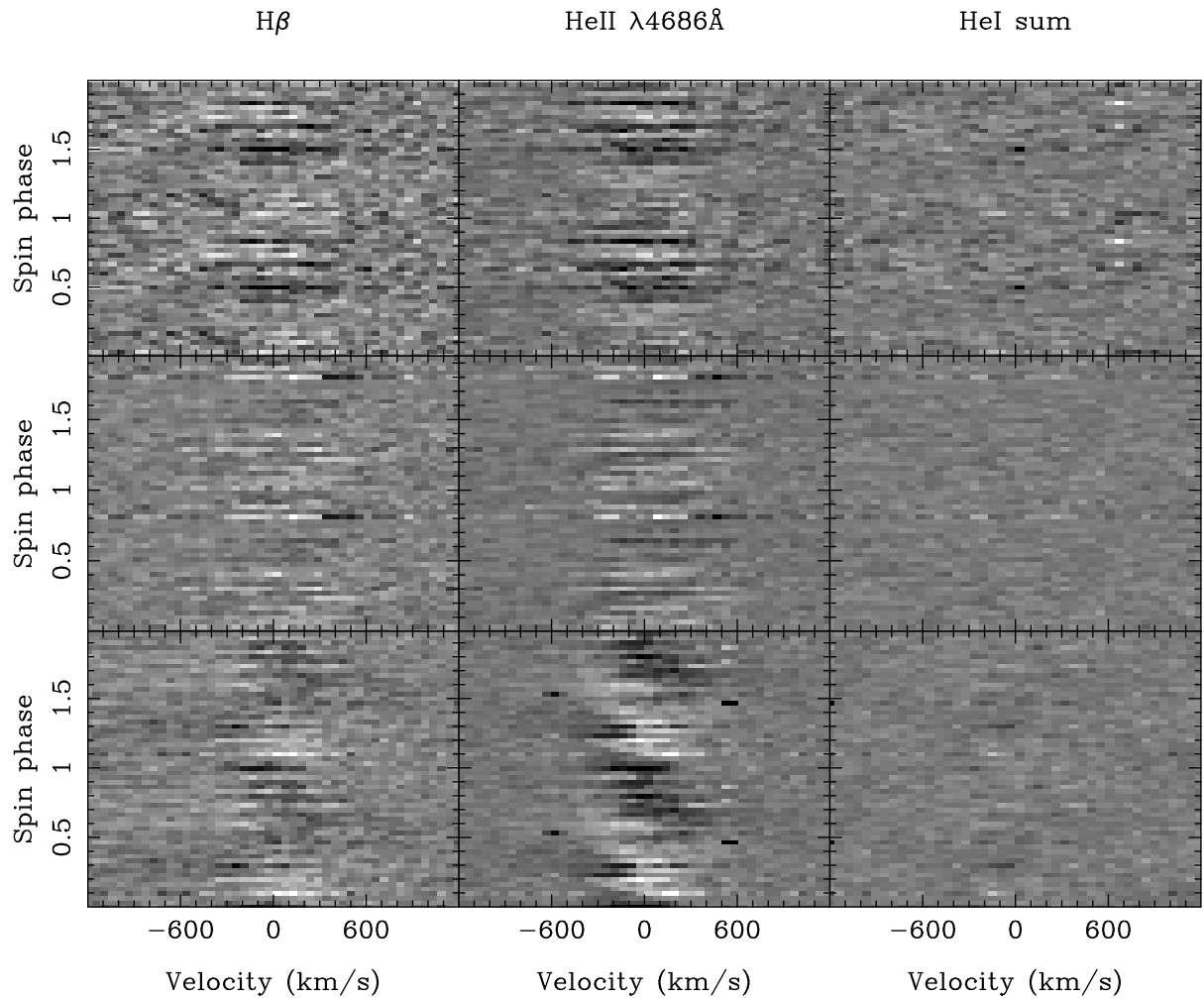}}
\noindent
\end{picture}
\vspace{140mm}
\caption{Trailed spectra of \hb, \heii\ and the average of He{\sc i} (as in
  Fig. 3) binned in the spin period. Note that the spin cycle is
  repeated twice. As in Fig.~3, each row represents a different night,
  the top row being the first night. Each image is on a separate
  linear intensity scale.}
\end{figure*}

The continuum-subtracted data are presented as time-series of selected
line profiles over each night of observation in Fig.~4.  At times
theses lines display double-peaked profiles.  This is often considered
a signature of accretion disc emission (Smak 1981), however the
velocity structure across the accretion flow is so complex in IPs that
this cannot be considered a conclusive detection of the accretion
disc.  The profiles are asymmetric where the peak apparently shifts
from the blue to the red and back again in the Balmer lines over the
observations.  This behaviour is reminiscent of emission from a
localised region in the system such as the bright spot where the
accretion stream strikes the outer rim of the disc, or an irradiated
region on the secondary star, but we note that the orbital phasing of
the observed \hb\ feature is inconsistent with both interpretations.
Moreover, the orbital phases at which we see these variations are
not those at which the hot spot and the irradiated face of the
secondary are best observed, i.e. phases 0.8 and 0.5 respectively.
The profile variations of the He{\sc i} and \heii\ lines are different
to those of \hb\ either because they originate from different
locations or are more sensitive to intervening absorption regions.

The most interesting variation occurs in the blue wings of all these
profiles.  First we note that the profiles are asymmetric about their
rest velocities, regardless of orbital phase, where each line has a
red bias. This shift is much larger than the systemic velocity of the
binary, measured from secondary star photospheric lines.  Secondly we
note that this asymmetry is periodic, at least in \hb\ and \heii, and
this period corresponds to the kilo-second QPOs we reported in paper
{\sc i}. The QPOs manifest in blue-shifted material and appear to be
the result of absorption either of the line source or the underlying
continuum.  We have presented trails of \hb\ and \heii\ against QPO
phase after subtracting the nightly average from each spectrum in
paper {\sc i}.

After the orbital period, the third likely signal present in these
trails is the 351\,s spin period of the white dwarf.  We attempted to
remove the orbital variations in the line profile by shifting out the
motion of the white dwarf according to the ephemeris and radial
velocity fit of CCF.  The QPO contribution was accounted for
approximately by combining the resulting spectra into 40 bins phased
over the QPO cycle and subtracting the spectrum in the bin nearest in
time from each individual spectrum.  Fig.~5 shows the resulting trails
of \hb, \heii \ and the sum of the He{\sc i} lines binned into 30
bins, over the spin period using Ishida's \etal (1992) ephemeris.  
The modulated signal is faint during the Feb 26 and 27, but clearly
present in the \hb\ and \heii\ profiles on Feb 28 extending out as far
as $\sim$\,1\,000\kmsec\ in the \hb\ profile. We see one modulation
per cycle with signal moving from the red peak to the blue. This is
reminiscent of the spin signal found in the trails of the IP
RX\,J0558+5353 (Still, Duck \& Marsh 1998), although in that case most
of the power occurred on the 1st harmonic of the spin period,
indicating accretion onto two poles of the primary star.  In the
current trails of \gk\ we see no evidence for power on the 1st
harmonic.  Harlaftis \& Horne (1998) postulate that the origin of this
spin pulsed emission is the region where the disc material is threaded
onto the magnetic field. Similar trails but for spectra obtained
during quiescence are presented by Reinsch (1994), where the spin
signal is also clear on the fundamental frequency in the Balmer and
\heii\ lines but not very strong in He {\sc i}.

\subsection{V/R ratios}

By measuring the fluxes under the continuum-subtracted blue wings
(from -1100 to 0 \kmsec) of the \hb\ and \heii\ lines and dividing
these by the flux under the red wings (from 0 to 1100 \kmsec) we
produce a time-series of V/R ratios. These are plotted in Fig.~6 for
the three nights of observation.  Kilo-second oscillations are
observed over all three nights.  A power search over the ratios was
performed using the Lomb-Scargle algorithm (Scargle 1982) and the QPO
periods found are listed in Table~5. The errors quoted are only an
estimate of the minimum error and depend on the frequency sampling. A
significance test (a variant of the randomisation Monte Carlo
technique, Linnell Nemec \& Nemec 1985) was run by iteratively
searching for periods after small shifts of the data had been
performed. After 1000 permutations we found that the periods present
in the V/R ratios are within the quoted errors with 95 per cent
confidence.  Note that the QPOs tend to shorter periods over the three
nights.

\begin{table}
\centering
\begin{minipage}{84mm}
\caption{QPO periods measured from the V/R ratios of \hb\ and \heii.}
\begin{center}
\begin{tabular}{llll}
1996 Feb&P$_{QPO}$&1996 Feb&P$_{QPO}$\\
\hline
\hb\ 26th& 6362 $\pm$ 3&\heii\ 26th& 6568 $\pm$ 3\\
\hb\ 27th& 5025 $\pm$ 2&\heii\ 27th& 5241 $\pm$ 2\\
\hb\ 28th& 3964 $\pm$ 1&\heii\ 28th& 4333 $\pm$ 1\\
\end{tabular}
\end{center}
\end{minipage}
\end{table}

\begin{figure*}
\begin{picture}(100,0)(10,20)
\put(0,0){\includegraphics{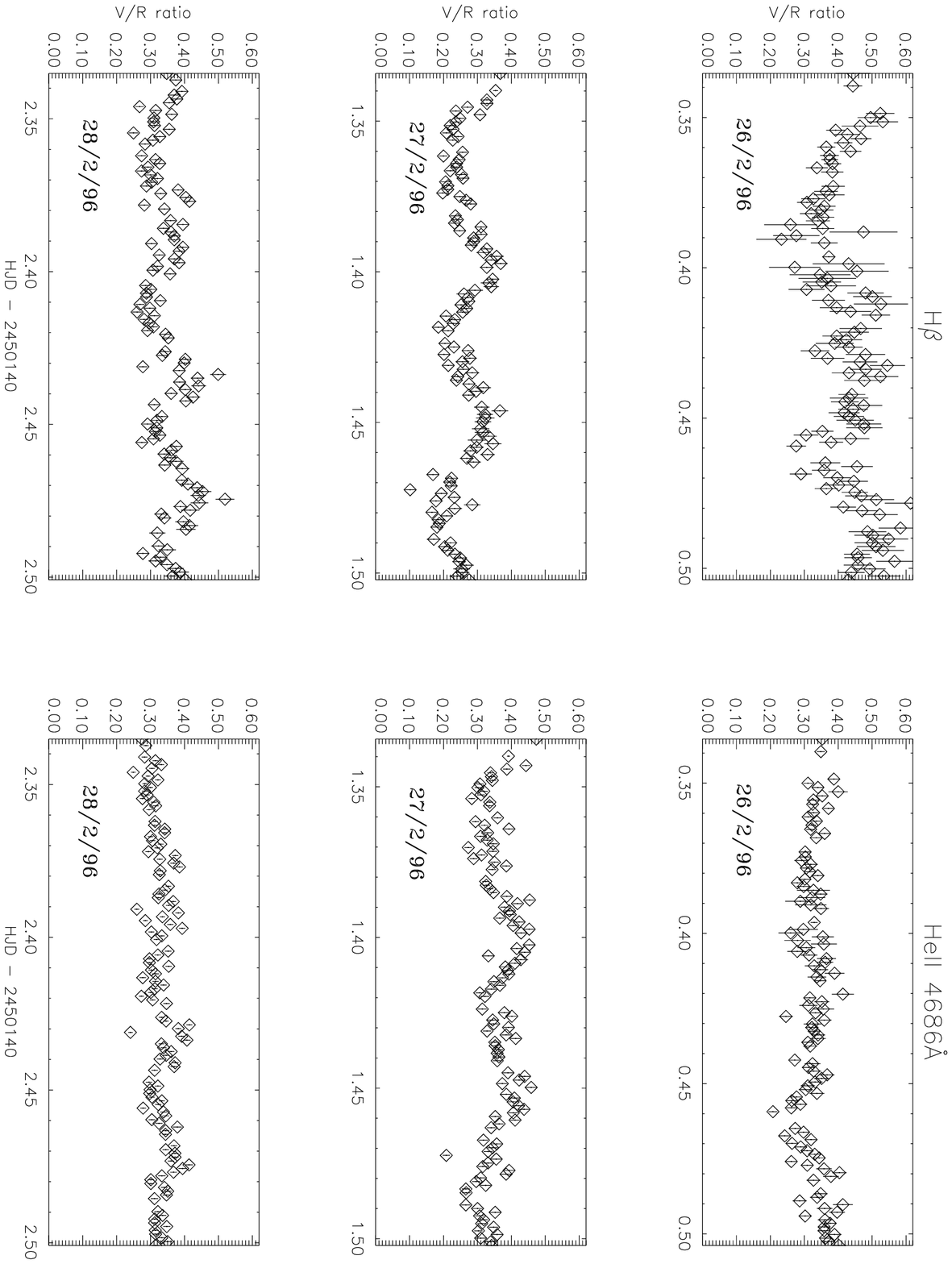}}
\noindent
\end{picture}
\vspace{140mm}
\caption{V/R ratio versus HJD for \hb\ and \heii. We find QPOs
  during all three nights.}
\end{figure*}

In order to determine the nature of higher-frequency variations we
created a coarse version of each V/R curve over 40 time bins on each
night, each bin approximating to one white dwarf spin cycle.  The QPO
signal was filtered out by subtracting the bin nearest in time from
each V/R measurement.  We searched for power in the modified V/R
ratios using the Lomb-Scargle algorithm and obtained the power spectra
plotted in Fig~7.

\begin{figure*}
\begin{picture}(100,0)(10,20)
\put(0,0){\includegraphics{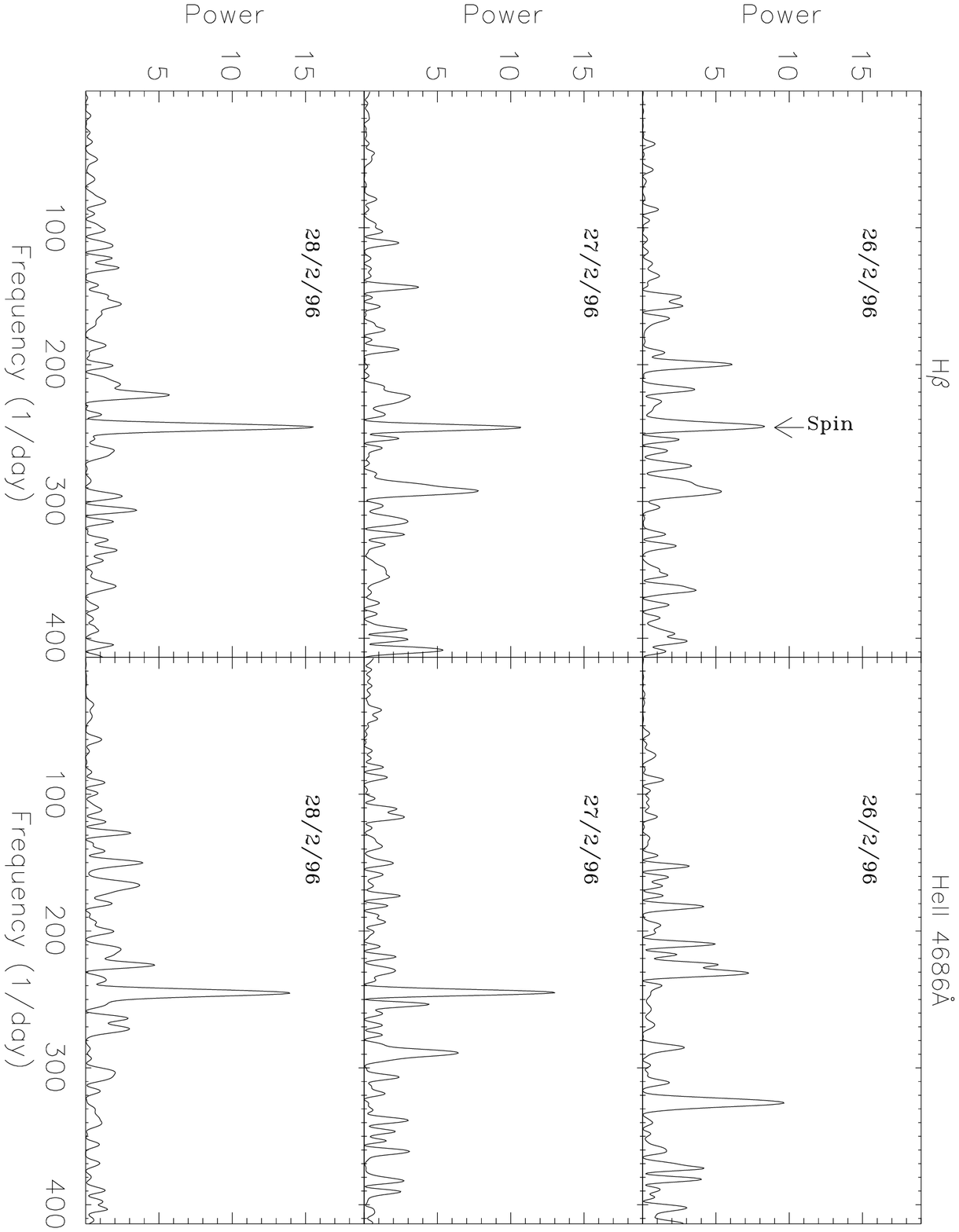}}
\noindent
\end{picture}
\vspace{120mm}
\caption{Power spectra obtained from the V/R ratios of \hb\ and \heii\ 
on 1996 Feb 26 to 28 after the QPO has been filtered out.}
\end{figure*}

\begin{figure*}
\begin{picture}(100,0)(10,20)
\put(0,0){\includegraphics{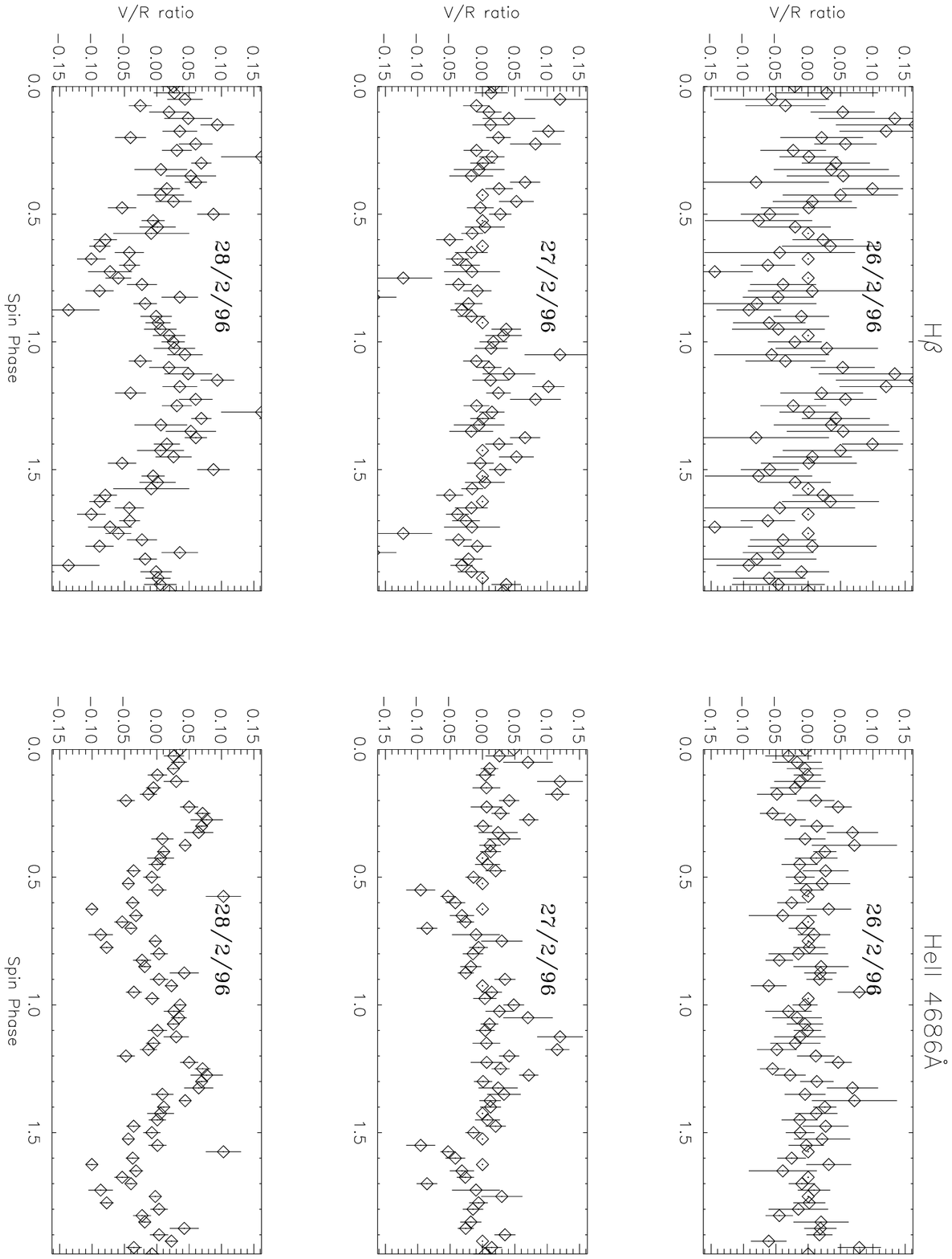}}
\noindent
\end{picture}
\vspace{140mm}
\caption{The QPO-filtered V/R ratios versus spin phase obtained
for \hb\ and \heii\ on 1996 Feb 26 to 28. The data are repeated over a
second spin cycle.}
\end{figure*}

Four of the spectra clearly show power at the white dwarf spin period
(246 cycles day$^{-1}$), but we cannot determine whether there is any
power on the 1st harmonic which occurs beyond the Nyquist limit.  We
folded the modified V/R ratio data using the spin ephemeris from
Ishida \etal (1992) using 40 bins and plot them in Fig.~8. The V/R
ratios show sinusoidal behaviour but not as clearly as in Garlick
\etal (1994) and Reinsch's (1994) quiescence data. The maximum in
these curves has previously been observed in quiescence at spin phase
0 rather than phase 0.25.

\subsection{Emission line and continuum fluxes}

We conducted a power search across the same regions of continuum
listed in Sec.~3.1 and the integrated flux over each emission line.
Power is present at kilo-second periods whose values decrease on
consecutive nights; see Table~6 and in Fig.~9 we present the power
spectra for the integrated fluxes of \hb, \heii\ and the continuum.
No significant power was found in the spin period during any of the
nights.  Garlick \etal (1994) and Reinsch (1994) find clear
modulations in the intensity of the Balmer lines with the spin period
during quiescence but no significant contributions at the spin period
in the continuum.

\begin{figure*}
\begin{picture}(100,0)(10,20)
\put(0,0){\includegraphics{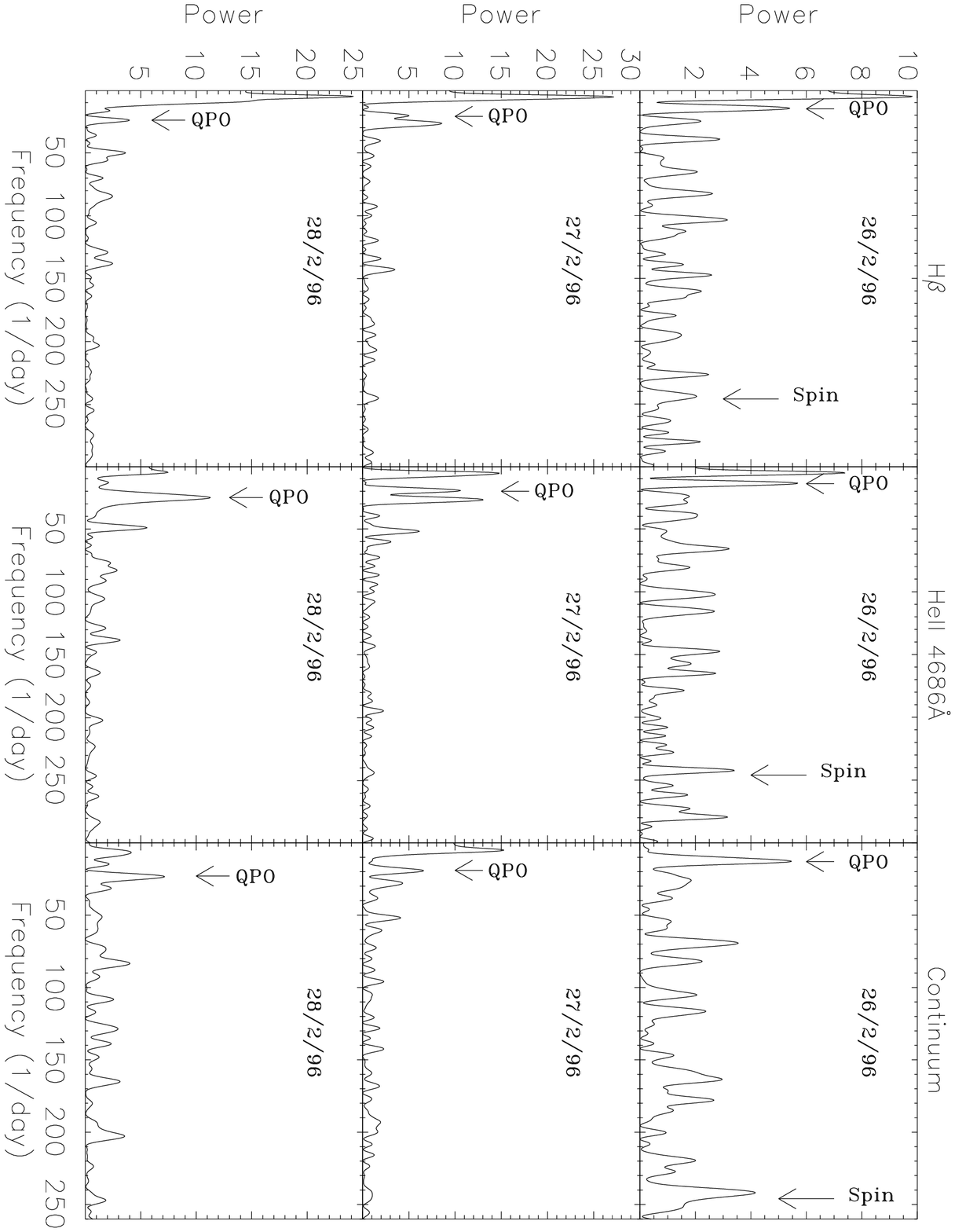}}
\noindent
\end{picture}
\vspace{140mm}
\caption{Power spectra for the integrated fluxes of \hb, \heii\ and
  the continuum.}
\end{figure*}

\begin{figure*}
\begin{picture}(0,0)(20,30)
\put(0,0){\includegraphics{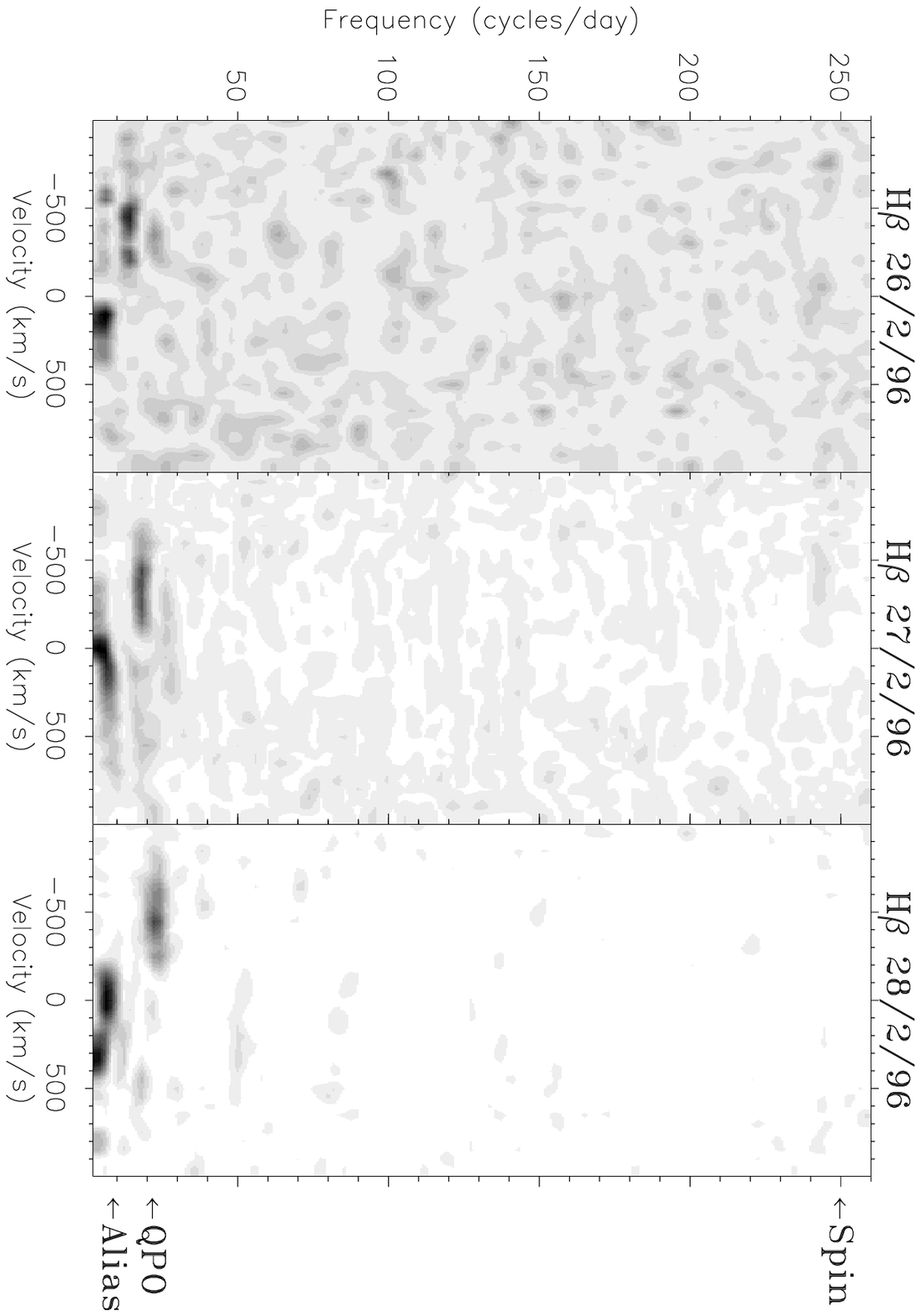}}
\put(0,0){\includegraphics{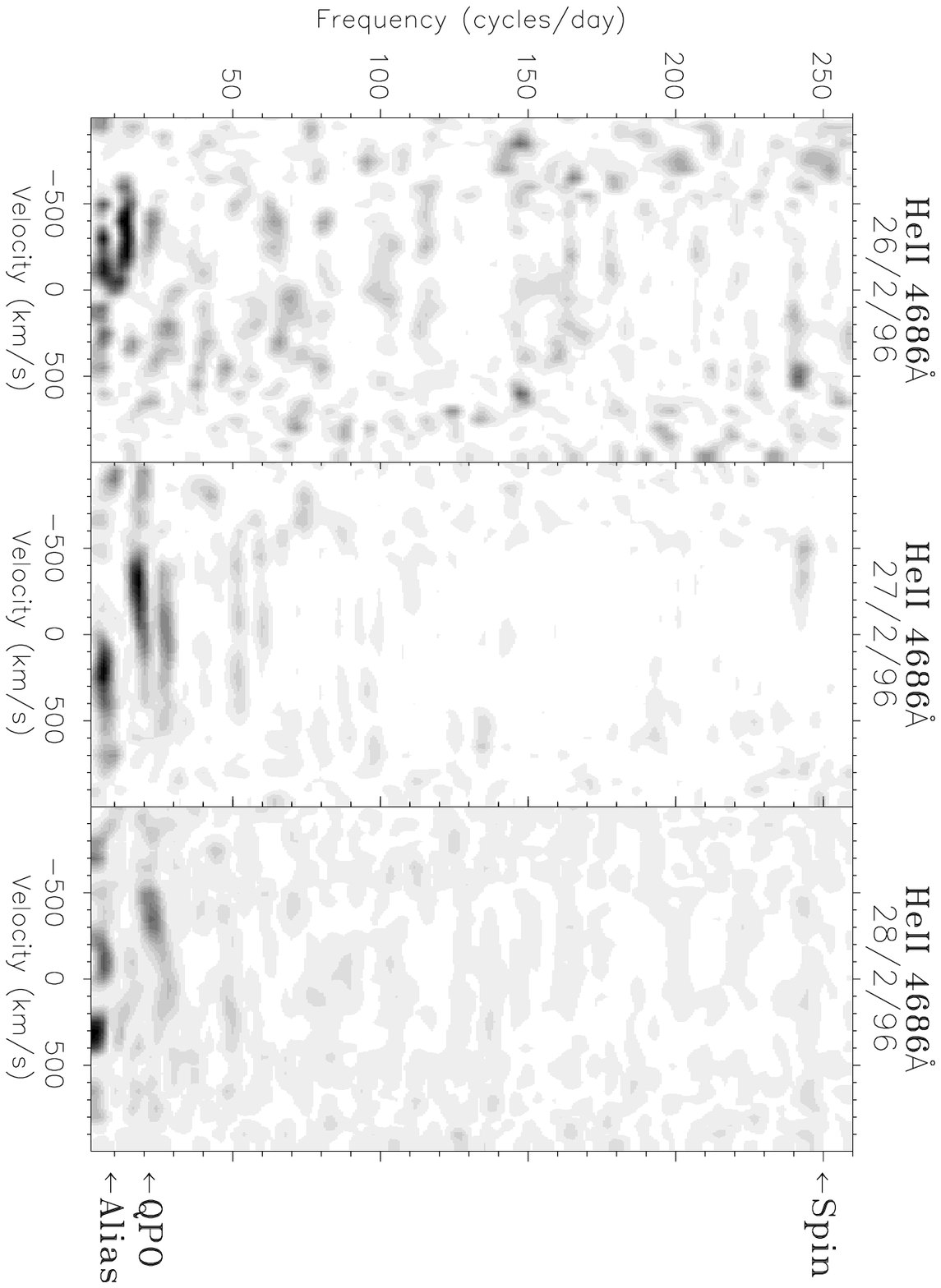}}
\noindent
\end{picture}
\vspace{175mm}
\caption{Power maps of frequency versus line velocity obtained from \hb\ 
  and \heii\ emission line fluxes of \gk.  We detect power at the
  4000--6000\,s frequency predominantly across the blue wings of these
  lines.  We detect further power at the white dwarf spin frequency
  ($\sim$~246 cycles/day).}
\end{figure*}

By means of a power search across radial velocity data, paper {\sc i}
determined that the characteristic velocities of the QPO signal were
intermediate between the orbital regime and the spin regime. This is
consistent with a signal origin in the inner disc or the threading
region between disc and magnetic curtain. However the search had no
means to discriminate between blue- and red-shifted material and
therefore could say little more about the QPO mechanism. In this paper
we conduct a similar power search but in the line fluxes across
discrete velocity bins of the emission profiles. The resulting power
maps for \hb\ and \heii\ are plotted in velocity--frequency space and
provided in Fig.~10, and Table~7 lists the QPO period on differentq
nights sampled at $-$400\kmsec\ and $-$600\kmsec\ in the line profile.
 
We find that power associated with the QPO is not symmetrically
distributed about the rest wavelength, but biased towards the blue
wing of each line, as we have previously noted in Sec.~3.3.  QPO power
extends from $-$500 to $-$1\,000\kmsec\ depending on the line and
night but the QPO does not appear to be a strong function of velocity
consistent with our results from paper {\sc i}.  We also find, the QPO
tending to longer frequencies with time, as we have already determined
from paper {\sc i} and the V/R ratio analysis in the current paper. We
discuss the significance of this result in Sec.~4.

We also  find power  on the  white dwarf spin  frequency extending  to
$\sim$\,1000\kmsec, although weaker than we  have found in the  radial
velocity analysis of paper {\sc i}.

\begin{table}
\centering
\begin{minipage}{84mm}
\caption{The QPO period determined from the integrated flux of \hb\
  and \heii\ and for the continuum the three nights of observations.
  The error given is the minimum error due to the frequency sampling.
  The third column gives an estimate of the confidence of the
  identification. This estimate has been obtained by using a
  randomisation Monte-Carlo Technique over 1000 permutations.}
\begin{center}
\begin{tabular}{lll}
  1996 Feb        & $P_{\mbox{\sevensize{QPO}}}$ (s) & Percentage \\
 & & of Confidence\\
\hline
  \hb\ 26th        & 5932 $\pm$ 2 & 89\\
  \hb\ 27th        & 4221 $\pm$ 1 \& 3246 $\pm$ 1 &  95 \& 95\\
  \hb\ 28th        & 3596 $\pm$ 1 & 93\\
  He{\sc ii}\ 26th & 6311 $\pm$ 2 & 90\\
  He{\sc ii}\ 27th & 4435 $\pm$ 1 \& 3228 $\pm$ 1 &95 \& 95\\
  He{\sc ii}\ 28th & 3447 $\pm$ 1 & 95\\
\hline
  Continuum 26th   & \multicolumn{1}{l}{ 6750 $\pm$ 26 } & 70\\
  Continuum 27th   & \multicolumn{1}{l}{ 4477 $\pm$ 12 } & 95\\
  Continuum 28th   & \multicolumn{1}{l}{ 3692 $\pm$ 9 }  & 94\\
\end{tabular}
\end{center}
\end{minipage}
\end{table}

\begin{table}
\centering
\begin{minipage}{84mm}
\caption{The QPO period determined from bins across the emission line
  profiles of flux of \hb\ and \heii\ on the three nights.  Errors
  are, as in Table~6, the estimates of the minimum error. All periods
  have been found to be significant to 95 per cent.}
\begin{center}
\begin{tabular}{lll}
  1996 Feb & $P_{\mbox{\sevensize{QPO}}}$ (s) & $P_{\mbox{\sevensize{QPO}}}$ (s) \\
  &at $-$400 km~s$^{-1}$ & at $-$600 km~s$^{-1}$ \\ \hline
  \hb\ 26th        & 6336 $\pm$ 2 & --- \\
  \hb\ 27th        & 4765 $\pm$ 1 & 4645 $\pm$ 1 \\
  \hb\ 28th        & 3766 $\pm$ 1 & 3637 $\pm$ 1 \\
  He{\sc ii}\ 26th & 6445 $\pm$ 2 & 6378 $\pm$ 2 \\
  He{\sc ii}\ 27th & 5912 $\pm$ 1 & --- \\
  He{\sc ii}\ 28th & 3725 $\pm$ 1 & 4038 $\pm$ 1 \\
\end{tabular}
\end{center}
\end{minipage}
\end{table}

\section{Discussion: QPO mechanisms}

In paper {\sc i}\ we found that the optical QPOs observed during the
1996 outburst have a velocity structure that is consistent with a
mechanism where dense blobs of gas orbit in the inner disc at a radius
determined by the impact of an overflowing gas stream (HL).  We
determined that the QPO is an approximately constant function of
velocity ruling out mechanisms involving radial or vertical
oscillations in the inner disc flow (Carroll \etal 1985), and that the
QPO tends to higher frequencies with time. In this paper we have shown
that the QPO is biased towards blue-shifted material and we discuss
this result in terms of the disc-overflow accretion model and the
alternative beat model proposed by WKO.

\subsection{The Disc-overflow accretion model}

This model was proposed by HL on the basis that a 5000\,s period is
consistent with the Keplerian period of blobs of material deposited in
the inner disc by the overflowing gas stream (Lubow \& Shu 1975; Lubow
1989; Hellier 1993; Armitage \& Livio 1996, 1998), and that the X-ray
hardness ratio measured as a function of QPO phase from the data
collected by WKO is consistent with the photo-electric absorption of
soft X-rays by cool intervening gas.  The optical counterpart to the
X-ray QPOs could either be direct reprocessing off the blobs or the
periodic reprocessing off material in the outer disc as the blobs
intermittently absorb the X-rays from the central object.

We have determined that, in velocity, the QPO ranges between the
expected rotational velocity of the outer disc and the overflow impact
site (paper {\sc i}). Although this distribution of QPO power is
consistent with the proposed mechanism, it is more difficult to
reconcile the blue-shifted bias of power in terms of the overflow
model (Fig.~10).

It is unlikely that the optical QPOs can be the result of reprocessing
off the disc unless there is an emission mechanism within the disc
which is extremely anisotropic. Similarly direct emission from the
blobs must also be anisotropic. In this case the cooling of shocked-
or viscous-heating gas within the blobs could cause the anisotropy
provided the blobs are orbiting faster than the surrounding disc
material. It is not clear why this should be the case.

\subsection{A disc-curtain beat mechanism}

\begin{figure*}
\begin{picture}(0,0)(20,30)
\put(0,0){\includegraphics{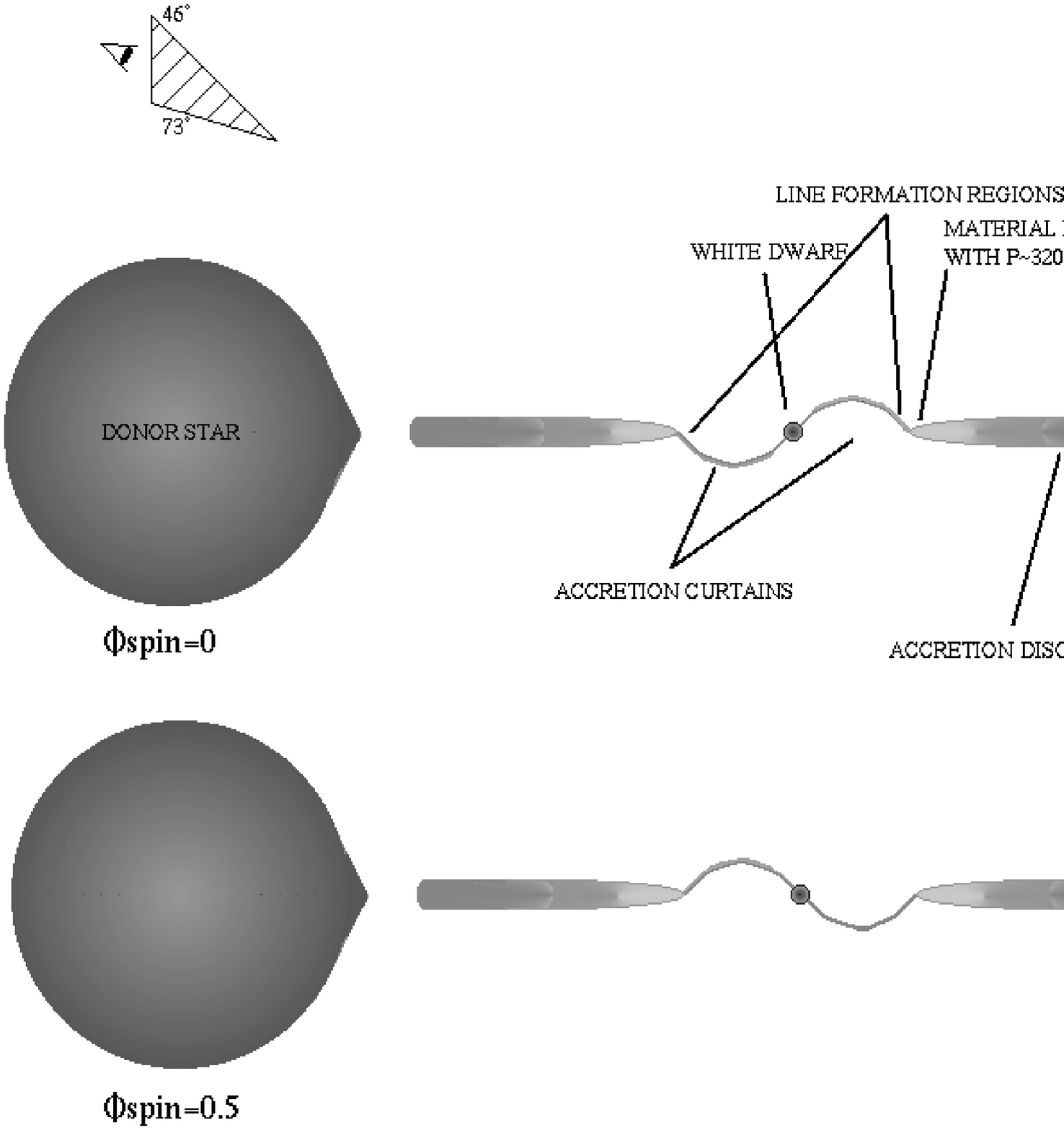}}
\noindent
\end{picture}
\vspace{130mm}
\caption{Schematics of the accretion flow in the intermediate polar
  \gk. The position of the observer corresponds to orbital phase zero.
  The two diagrams are slices through a vertical plane across the
  binary line of centres, separated in time by half a spin cycle.}
\end{figure*}

WKO proposed a QPO mechanism where the observed periods are the beats
between the spin frequency of the white dwarf and the Keplerian
frequency of dense blobs of gas orbiting at the inner rim of the disc
(Alpar \& Shaham 1985 a, b).  Each time an accretion curtain sweeps
over a blob we observe an increase in column density across the
curtain, providing a cool absorbing body for the X-ray emission. The
prediction follows that the orbital period of the inner disc is either
$\sim$\,320\,s or $\sim$\,380\,s. We investigate whether this model
can explain the observed bias in the QPOs across the optical emission
lines. Two schematics of the accretion flow in the binary are depicted
in Fig.~11.  We take the orbital inclination to be consistent with
$46^{\circ} < i < 72^{\circ}$ (Reinsch 1994).  We have assumed that
the magnetic axis of the white dwarf is misaligned with the rotational
axis of the system by $45^{\circ}$, although its true inclination is
unknown.

In Sec.~3.3 we determined that similar to the X-ray QPOs, the optical
counterpart in the emission lines are the result of absorption. We
consider two mechanisms which modulate the line emission by absorption
over the WKO beat cycle. First we consider self-absorption of line
flux from the curtain just above each threading region. Our
observation that QPO signal does not extend to as large velocities as
the spin signal provides some justification for this assumption. We
require the absorption profile to be saturated such that line strength
is a function of both column density through the curtain and the
velocity gradient across the flow (see e.g.\ Horne \& Marsh 1986). At
spin phase equal zero, \phispin\ = 0 the upper accretion curtain lies
behind the white dwarf and material flowing along that curtain is
blue-shifted.  Conversely material in the curtain feeding the lower
pole is red-shifted. We should observe a blue shifted QPO signal when
the blob sweeps through the first threading region every few thousand
seconds, increasing the column density along the approaching curtain,
but not a red-shifted signal because the second threading region is
obscured by the inner accretion disc.  The result is a blue bias in
the QPO signal across the emission lines at this spin phase. However
at \phispin\ = 0.5 the curtain geometry has rotated by 180$^{\circ}$
and both curtains are equally visible.  But in this configuration the
velocity gradient across the line forming regions is small compared to
the \phispin\ = 0 case and consequently the amount of absorption
across the line profile is smaller. In this way the blue bias in the
signal is conserved over the beat cycle.

An equally plausible alternative is that the absorption is of
continuum light from the accretion disc behind the white dwarf. As
before, the column density along the accretion curtains is modulated
on the beat cycle as a blob sweeps through the threading regions.
This provides a kilo-second QPO by periodic absorption which occurs
when the upper curtain is back-illuminated by the disc at \phispin\ =
0. Since there is no back-illuminating source for the lower curtain,
or for the upper curtain when it is red-shifted, this mechanism
provides a natural blue bias to the QPO signal.

The beat model explains the long-timescale QPOs from \gk\ using the
pre-existing models of QPO generation. In these models the driving
mechanism has a timescale of a few hundred seconds, as observed in the
rest of the dwarf nova class of objects. The extra ingredient for \gk\ 
is provided by its properties as both a dwarf nova and an intermediate
polar, where the QPO beats with the accretion curtains which thread
the disc onto the rapidly spinning white dwarf to provide the observed
kilo-second periods. In the above discussion we have considered the
QPO in terms of a blob mechanism but the alternative models of
radially-oscillating acoustic waves in the inner accretion disc work
equally well. (Okuda \etal 1992; Godon 1995). Consequently we do not
require a new physical explanation of QPOs to explain the phenomenon
in \gk.

\subsection{QPOs or DNOs?}

QPOs are present in dwarf novae during quiescence and outburst.
However the kilo-second QPOs in \gk\ have to date only been found when
the system is in outburst (Reinsch (1994) claims a tentative
kilo-second detection in optical photometry but provides no evidence).
This behaviour is more typical of another class of oscillations -- the
dwarf nova oscillations (DNOs), which occur on timescales of tens
rather than a few hundred seconds (Robinson and Nather 1979).  A
characteristic of DNOs is that they tend to shorter periods as a
system approaches the peak of its outburst (Patterson 1981), similar
to what we have found in the current observations.  Given that the
beat model is correct, the oscillations found in \gk\ show
characteristics of both QPOs and DNOs.  The timescales are consistent
with QPOs, whereas their behaviour is more comparable to DNOs.

The timescales of DNOs suggest they are driven at the inner edge of an
accretion disc (see Warner 1995 for a review in DNOs), and since \gk\ 
is an unusual dwarf nova in that the inner disc is truncated by the
white dwarf field, it seems plausible that a DNO mechanism in this
system would work on a larger timescale. If we are observing DNOs, the
implication of the period increase over the three nights for the beat
model is that the inner disc must be orbiting at 320\,s rather than
380\,s. In the latter case a decrease in the period of the driving DNO
from the disc will result in an increase in the observed beat period.
We do not find modulations in our data with either a 320\,s or 380\,s
period. Tentative detections of signal at 390 $\pm$ 20\,s and 410
$\pm$ 13\,s have been claimed by Mazeh \etal (1985) from optical
photometry during the 1983 outburst while Patterson (1981) reports a
detection at 380 $\pm$ 20\,s but never presented the result. If
present during the quiescent state, a 380\,s oscillation would suggest
a QPOs classification. There have been no reports of a 320\,s period
in the literature.

\section{Conclusions}
We obtain new values for the systemic velocity and the velocity
semi-amplitude of the secondary in agreement with previous authors. We
conclude that there is no evidence for increased heating over the
inner face of the donor star during this stage of the outburst.  We
find spin modulations in the V/R ratios of the lines but only
tentatively in their integrated fluxes or the continuum. Spin power
resolved across the line profiles extends to velocities of 1000\kmsec,
a large fraction of the freefall velocity of the central object.

The detection of kilo-second QPOs across the optical emission line
profiles of \gk\ have provided an opportunity to test the mechanism
behind the unique long-timescale QPOs in this object, which are an
order of magnitude longer than QPOs normally observed in
disc-accreting cataclysmic variables.  We have rejected the model of
HL which considers the direct effects of blobs orbiting at the
Keplerian frequency of the annulus associated with a disc-overflow
impact site.  Our favoured models consider the long QPO period to be
the consequence of beating between more typical timescale QPOs or DNOs
of $\sim$~300--400\,s with the magnetic accretion curtain spinning
with the white dwarf.  Therefore we do not require a new model to
explain these long timescales -- the long oscillations are merely a
consequence of the magnetic nature of the binary.

\section*{ACKNOWLEDGEMENTS}

We thank Janet Wood and John Lockley for obtaining the spectra of the
K-type templates at the McDonald Observatory.  MDS was supported by
PPARC grant K46019.  PDR acknowledges the support of the Nuffield
Foundation via a grant to newly qualified lectures in science to
assist collaborative research. The reduction and analysis of the data
were carried out on the Sussex node of the STARLINK network. We thank
Tom Marsh for providing his reduction software. LM also wishes to
thank R. I. Hynes for useful discussion. The Isaac Newton Telescope is
operated on the island of La Palma by the Isaac Newton Group in the
Spanish Observatorio del Roque de los Muchachos of the Instituto de
Astrof\'{\i}sica de Canarias.

\label{lastpage}

\end{document}